\documentclass[aps, prl, twocolumn,superscriptaddress,showpacs]{revtex4}
\usepackage{graphicx}
\usepackage{epsfig}
\usepackage{amssymb}

\begin{document}

\title{Spatial Coherence of a Polariton Condensate}
\author{Hui Deng}
%\email{hdeng2@caltech.edu}
\altaffiliation{Current Address: Norman Bridge Laboratory of Physics
12-33, California Institute of Technology, Pasadena, California
91125, USA}\affiliation{Edward L. Ginzton Laboratory, Stanford
University, Stanford, California 94305, USA}
\author{Glenn S. Solomon}
\affiliation{National Institute of Standards and Technology, Physics
Laboratory, 100 Bureau Drive, MS 8423, Gaithersburg, MD 20899, USA}
\author{Rudolf Hey}
\author{Klaus H. Ploog}
\affiliation{Paul-Drude-Institut f\"{u}r Festk\"{o}rperelektronik,
Hausvogteiplatz 5-7, D-10117 Berlin, Germany}
\author{Yoshihisa Yamamoto}
\affiliation{Edward L. Ginzton Laboratory, Stanford University,
Stanford, California 94305, USA} \affiliation{National Institute of
Informatics, Tokyo, Japan}

\begin{abstract}
We perform a Young's double-slit experiment to study the spatial
coherence properties of a two-dimensional dynamic condensate of
semiconductor microcavity polaritons. The coherence length of the
system is measured as a function of the pump rate, which confirms a
spontaneous build-up of macroscopic coherence in the condensed
phase. An independent measurement reveals that the position and
momentum uncertainty product of the condensate is close to the
Heisenberg limit. An experimental realization of such a minimum
uncertainty wavepacket of the polariton condensate opens a door to
coherent matter-wave phenomena such as Josephson oscillation,
superfluidity, and solitons in solid state condensate systems.
\end{abstract}
%\pacs{
%71.36.+c, % Polaritons (including photon-phonon and photon-magnon interactions)
%42.50.-p,  % Quantum optics
%78.47.+p,  % Time-resolved optical spectroscopies and other ultrafast optical measurements in condensed matter
%78.67.-n, %Optical properties of low-dimensional, mesoscopic, and nanoscale materials and structures
\maketitle

Simple, yet profoundly connected to the foundation of quantum
physics, the Young's double-slit experiment has been a benchmark
demonstration of macroscopic spatial coherence -- off-diagonal long
range order (ODLRO) of a macroscopic number of particles
\cite{Penrose_PR56} -- in Bose-Einstein Condensation (BEC) of cold
atoms \cite{Andrews1997b,Bloch2000a,Hadzibabic2006a}. Recently, a
similar phase transition has been reported for the lower branch of
exciton-polaritons (LPs) in planar semiconductor microcavities
\cite{Dang_PRL98,Senellart_PRL99,Robin_PRB02,Deng_SCI02,Deng_PNAS03,Dang_PRL05,Deveaud_NA06,Deng_PRL06},
and supporting theoretical frameworks have been developed
\cite{Littlewood_JP04,Keeling_SST07,Malpuech_SST03,Kavokin_PRL04,Cao_PRB04,Doan_PRB05,Sarchi_arxiv06,Baas_PRL06}.
Interestingly, LPs are free particles in a two dimensional (2D)
system where genuine BEC exists only at zero temperature in the
thermodynamic limit \cite{Mermin1966a,Hohenberg1967a}. A quasi-BEC
can be defined for a 2D system of a finite size if a macroscopic
number of particles occupy a single ground state and if an ODLRO is
established throughout the system
\cite{Kleppner_PRA91,Lauwers_JPa03}. Yet in the LP experiments to
date, the system size is ambiguously defined by the spot size of the
pump laser, and there is no quantitative study of the relation
between the size and the coherence length of a condensate
\cite{Dang_PRL05,Deveaud_NA06}. In this work, we perform a Young's
double slit experiment on a LP gas to measure its spatial coherence
properties across the phase transition, and compare the measured
coherence length with the condensate size. We also measure the
position-momentum uncertainty product of the condensate and compare
it to the Heisenberg limit.

\begin{figure}[phtb]
     \includegraphics[width=3in,height=2in]{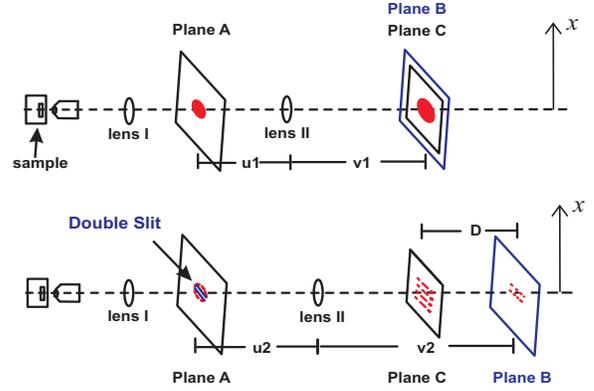}
     \caption{A sketch of the double-slit experiment setup. Upper: the LP spatial distribution is imaged to a CCD
at plane \textit{C}. Lower: a double-slit is inserted at plane
\textit{A} and imaged by lens II to a virtual plane \textit{B}. The
CCD at plane \textit{C}, a distance D from plane \textit{B},
captures the double-slit interference pattern.} \label{fig:setup}
\end{figure}
%\section{experimental setup}
A sketch of the setup is shown in Fig.~1. The microcavity sample is
first magnified by a factor of 37.5 and imaged to a plane
\textit{A}, which is in turn imaged by a lens II to a
charge-coupled-device (CCD) at plane C for measurement of spatial
distribution. For the double-slit experiment, we insert a pair of
rectangular slits at plane \textit{A}, and move the lens II such
that the image of plane \textit{A} (denoted by plane \textit{B}) is
a distance $D$ behind plane \textit{C}. Effectively, we observe on
the CCD the interference pattern of the LP emission passing through
the double-slit. In our experiment, $D=6.7$~cm, the width of the
slit image at plane \textit{B} is $\delta=53~\mu$m, and the average
wavelength of the LP emission is $\lambda\sim 778.5$~nm.
Correspondingly, the Fresnel number $\frac{\delta^2}{D\lambda}=0.05
\ll 1$, thus the far-field condition is satisfied at plane
\textit{C}. When mapped onto the sample surface, the slit width seen
by the LPs is $\Delta r\approx 0.5~\mu $m, which is less than the
intrinsic coherence length $\xi_0\sim 1~\mu$m of a single LP
\cite{note:xi0}, and much less than the LP system size of $5-
10\mu$m. Hence neglecting the variation in LP distribution within
each slit, we obtain the intensity distribution on the CCD camera
\cite{wolf_book}:
\begin{eqnarray}
I(x) &=& I_{1}(x) + I_{2}(x)+\nonumber\\
&&g^{(1)}(|r_1-r_2|)\cdot
2\sqrt{I_{1}(x)I_{2}(x)}\cos (\phi(x)+\phi_{12}),\nonumber\\
I_i(x)&=& |E (r_i)|^2\cdot \textrm{sinc}(\frac{x-x_0\pm  d/2}{X})\nonumber\\
\phi(x) &=& \frac{2(x-x_0)}{X_c},\quad X = \frac{2D}{k_{tot}\delta},
\quad X_c=\frac{2D}{k_{tot}d}.\label{eq:I_x}
\end{eqnarray}
The subscript $i=1,2$ denotes the slit number. $r_i$ are the
x-coordinates of the slits on plane \textit{B}
(Fig.~\ref{fig:setup}), $x$ is the x-coordinate on plane \textit{C},
$x_0$ is the center of the double-slit on plane \textit{C}, $E(r_i)$
is the LP field amplitude at slit $i$, $k_{tot}$ is the free space
average wavenumber of the LP emission, and $d$ is the separation
between the images of the two slits at plane \textit{B}. $I_i$ is
the intensity distribution if only slit $i$ is open. $\phi_{12}$ is
a fixed phase difference of LPs between the two slits. $\phi(x)$ is
a varying phase close to the path length difference from the two
slits, giving rise to a cosine modulation on the far-field intensity
distribution. After proper normalization, the amplitude of the
cosine modulation equals the first order coherence function
$g^{(1)}(r=|r_1-r_2|)$. Using six sets of double-slit with varying
slit separations $r$, we measured $g^{(1)}(r)$ from $1.3~\mu$m,
close to the intrinsic coherence length of a single LP, up to
$8~\mu$m, close to the LP system size. By varying the pumping
intensity, we studied the characteristics of $g^{(1)}(r)$ across the
phase transition.

The sample we investigated has a $\lambda/2$ GaAs cavity sandwiched
between Ga$_{0.865}$Al$_{0.135}$As/AlAs distributed Bragg
reflectors. Three stacks of quantum wells (QWs) are placed at the
central three antinodes of the microcavity, each stack consisting of
four 6.8~nm-wide GaAs QWs separated by 2.7~nm-wide AlAs barriers. We
pump the sample with linearly polarized pico-second mode-locked
Ti-Sapphire laser. At an incidence angle of $50^{\circ}$ from the
sample growth direction, the laser is resonant with the exciton-like
LP modes. The sample is kept at $T_{lattice}\approx 4$~K. The
cavity-photon energy at zero in-plane wavenumber is $\sim 7$~meV
above the bare QW exciton resonance, corresponding to an optimal
detuning for thermal-equilibrium condensation of the LPs
\cite{Deng_PRL06}. The threshold pumping density is $P_{th}\sim
300$~W/cm$^2$ \cite{note:threshold}.

\begin{figure}[phtb]
     \includegraphics[width=2.8in,height=4in]{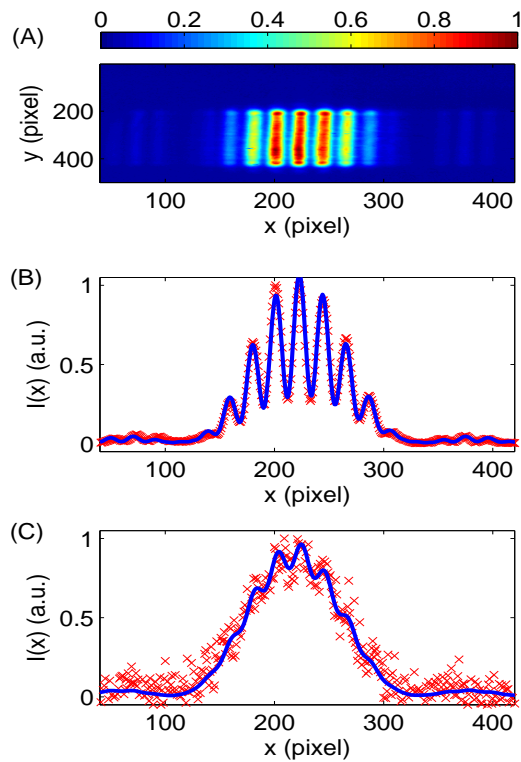}
     \caption{\textbf{(A)} Raw image of the interference pattern,
     slit separation $r=2.7~\mu$m, $P/P_{th}=7$. \textbf{(B)} Measured
     (symbols) intensity I(x) for $r=2.7~\mu$m, $P/P_{th}$=6.7 and fitting
     by Eq.~\ref{eq:I_x}. Fitted $g^{(1)}(r)=0.560\pm 0.006$.
     \textbf{(C)} Same as (B), for $r=2.7~\mu$m, $P/P_{th}=0.5$ and
     $g^{(1)}(r)=0.09\pm 0.02$.} \label{fig:img}
\end{figure}
%\section{Images}
A typical interference pattern observed at a pump rate above the
condensation threshold is shown in Fig.~\ref{fig:img}A. Distinct
interference fringes are readily observed imposed on a sinc function
distribution. To obtain $g^{(1)}(r)$, we integrate over a narrow
strip along the y-axis and fit it with Eq.~\ref{eq:I_x}. $g^{(1)}$,
$\phi_{12}$, $|E_1|$ and $|E_2|$ are free fitting parameters, while
$x_0$, $d$, $X$ and $X_c$ are estimated from experimental parameters
with a 10$\%$ allowed variation. As shown in Fig.~\ref{fig:img}B,
Eq.~\ref{eq:I_x} fits the data very well for $P>P_{th}$. At
$P<P_{th}$, the interference patterns are barely observable or
non-existing; one example is given in Fig.~\ref{fig:img}C.

\begin{figure}[phtb]
     \includegraphics[width=2.8in,height=4.2in]{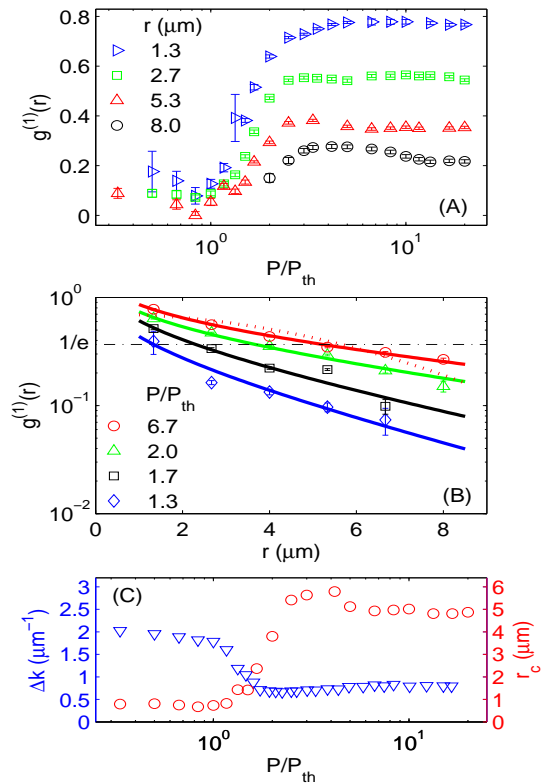}
     \caption{\textbf{(A)} $g^{(1)}(r)$ vs. $P/P_{th}$ for different
slit separations $r$ as labeled in the figure. \textbf{(B)}
$g^{(1)}(r)$ vs. $r$ for different pump rates $P/P_{th}$ as given in
the legend. The symbols are measured $g^{(1)}(r)$. The solid lines
are fittings by Eq.~\ref{eq:Kr}. % with, from top to bottom, $\Lambda_c$=20, 16.7, 11.7, 9.2~$\mu$m.
The dashed line at $P/P_{th}=7$ is a fitting by Eq.~\ref{eq:exp2}.
The dash-dotted line marks where $g^{(1)}(r)=1/e$. \textbf{(C)}
$r_c$, the $1/e$ decay length of $g^{(1)}(r)$, vs. $P/P_{th}$
(circles) and $\Delta k$ vs. $P/P_{th}$ (triangles).
}% with $\Lambda_c = 8~\mu$m.}
\label{fig:fc_ps}
\end{figure}

% g1 vs. p, for varying r
In Fig.~\ref{fig:fc_ps}A we show the increase of $g^{(1)}(r)$ with
$P/P_{th}$ for a few slit separations $r$. A salient feature in
Fig.~\ref{fig:fc_ps}A is that, there is a jump in $g^{(1)}(r)$ when
the pump rate is increased above a condensation threshold, even at
$r$ up to the laser pump spot size of $\sim 8~\mu$m. When the pump
spot size is increased to $\sim 20~\mu$m, $g^{(1)}(r)>0.3$ was also
observed up to $r=20~\mu$m, limited again by the pump spot size
\cite{Lai_spotsize}. This demonstrates the sudden appearance of
macroscopic coherence above the condensation threshold. Another
feature is that, the increase of $g^{(1)}(r)$ with $P/P_{th}$ is
slower at larger $r$. This shows that the macroscopic coherence is
built up gradually throughout the pump beam spot size when the phase
space density of LPs is increased.

% g1 vs. r, fit by proper models
To study the spatial coherence properties quantitatively, we plot in
Fig.~\ref{fig:fc_ps}B how $g^{(1)}(r)$ decays with $r$, and define a
coherence length $r_c$ as $g^{(1)}(r_c)=1/e$. % \cite{note:rc}.
The normalized pump power dependence of $r_c$ is shown in
Fig.~\ref{fig:fc_ps}C. As a reference, for a classical
Maxwell-Boltzmann (MB) gas in thermal equilibrium:
\begin{eqnarray}
g^{(1)}(r)&=&e^{-\pi r^2/\Lambda_T^2},\:\: r_c =
\frac{\Lambda_T}{\sqrt{\pi}} = \sqrt{\frac{2\pi\hbar^2}{mk_B T}}.
\label{eq:exp2}
\end{eqnarray}
Here $\Lambda_T$ is the thermal de Broglie wavelength, $m$ is the
mass of the particles and $T$ is temperature. For the current system
at $T=4$~K, $\Lambda_T/\sqrt{\pi}\approx 1.9~\mu$m.

Below condensation threshold density, we measured a
$r_c<\Lambda_T/\sqrt{\pi}$, since the system is far from thermal
equilibrium. In fact, the finite $g^{(1)}(r)\sim 0.15$ at
$r=1.3~\mu$m is consistent with the intrinsic coherence length
$\xi_0\approx 1~\mu$m of a single LP due to its finite lifetime
\cite{note:xi0}: $g^{(1)}_0(r) = \exp(-\frac{r^2}{\xi_0^2}) = 0.19$.
Above condensation threshold, we found $r_c\gg\Lambda_T/\sqrt{\pi}$.
Fitting of the data with Eq.~\ref{eq:exp2} with temperature $T$ as a
free parameter also fails to describe the data (dashed lines in
Fig.~\ref{fig:fc_ps}B).

Since $g^{(1)}(r)$ is the Fourier transform of the momentum
distribution $f(k)$, we resort to the actual momentum distribution
of the system. It was found that above the condensation threshold,
the LPs become highly degenerate in the states with the lowest
kinetic energies (e.g., Fig.~2 in Ref.~\cite{Deng_PRL06}). Their
momentum distribution deviates from the MB distribution, but follows
well the Bose-Einstein (BE) distribution with chemical potential
$|\mu|\ll k_B T$. In this quantum degenerate limit, $f(E\sim k_B T)
= [ exp(\frac{E-\mu}{k_BT})-1]^{-1}\approx (e-1)^{-1} \ll
f(0)\approx k_BT/\mu $, most of the emission comes from LPs with
$E\sim 0$. Hence we can obtain the following approximate form of
$g^{(1)}(r)$:
\begin{eqnarray}
g^{(1)}(r) &\propto& \mathcal{F}^{(2D)}(f(k )g(k ))\propto \mathcal{H}(f(k ))\nonumber\\
%&\sim& \mathfrak{H}(\[ exp(\frac{E(k )-\mu}{k_BT})-1\]^{-1} )\\
&\approx & \mathcal{H}( \frac{k_BT}{E(k )-\mu} ) \propto
K_0(r\frac{\sqrt{4\pi|\mu|/k_B T}}{\Lambda_T}).
 \label{eq:Kr}
\end{eqnarray}
Here $k $ is the LP's in-plane wavenumber, $g(k)$ is the constant
momentum density of state, $\mathcal{F}^{(2D)}(f(k))$ and
$\mathcal{H}(f(k ))$ denote the 2D Fourier and Hankel transform of
$f(k )$, respectively. As shown in Fig.~\ref{fig:fc_ps}B (solid
lines), $K_0(x)$, the modified Bessel function of the first kind,
fits very well the measured $g^{(1)}(r)$ for $P/P_{th}>1$.

If we extrapolate the $e^{-\pi}$ decay length of $g^{(1)}(r)$ from
the fitting and compare it to $\Lambda_T$ at 4~K, it is
$2.5\Lambda_T$ at $P=1.2P_{th}$, and $7.9\Lambda_T$ at
$P=6.7P_{th}$.

In Fig.~\ref{fig:fc_ps}C, we compare the pump rate dependence of
$r_c$ and $\Delta k $, the $1/e$ width of the measured first-order
coherence function $g^{(1)}(r)$ and the measured momentum
distribution function $f(k)$, respectively. At pump rates lower than
the condensation threshold, $r_c$ is $\sim 1~\mu$m, limited by the
intrinsic coherence length $\xi_0$ of a single LP, while $\Delta k$
is $\sim~2~\mu$m$^{-1}$ due to the slow energy relaxation dynamics
of the LPs. The product $r_c\cdot\Delta k$ is close to 2, the value
expected for a thermal MB distribution. When $P$ increases toward
$P_{th}$, more injected LPs relax to the lower energy states and
$\Delta k$ gradually narrows, but $r_c$ is still limited by $\xi_0$,
hence a decrease in $r_c\cdot\Delta k$. Once above the threshold,
there is a sudden increase of $r_c$ by more than five fold up to
$\sim 6~\mu$m, which manifests the spontaneous build-up of a global
phase among the LPs due to the phase coherent stimulated scattering
of LPs into the ground state. Correspondingly, $\Delta k$ is reduced
by about four fold since the LPs form a quantum degenerate Bose gas.
Further increasing the pump rates, $r_c$ decrease slightly while the
momentum distribution is broadened, potentially because stronger
LP-LP scattering at high densities introduces condensate dephasing
%(self-phase modulation)
\cite{Sarchi_arxiv06}.

% spatial spot size
Finally, it is instructive to compare $\Delta k$ with the measured
condensate size \cite{Deng_PNAS03}. We have consistently observed an
abnormally slow increase of the condensate size in comparison to the
spot size of a photon laser based on electron-hole pairs (Fig.~6C in
Ref.~\cite{Deng_PNAS03}).

\begin{figure}[phtb]
     \includegraphics[width=2.8in,height=3in]{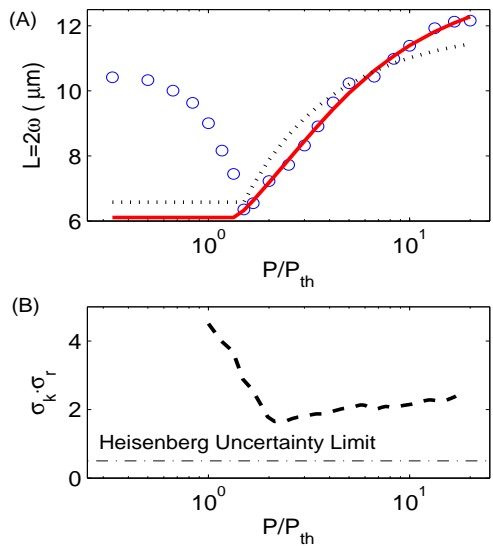}
     \caption{\textbf{(A)} System size $2\omega$ vs. pump rates $P/P_{th}$.
    Symbols are the data. The solid line is a fitting by
    Eq.~\ref{eq:omega_pol2}, with $\omega_p=13.3\pm 0.1~\mu$m, $\omega_c=6.1\pm 0.1~\mu$m. The dotted
    line is a fitting by Eq.~\ref{eq:omega_ph} for comparison, with
    $\omega_p=11.9\pm 0.2~\mu$m. \textbf{(B)} The position and momentum uncertainty
    product $\sigma_k \cdot \sigma_{r}$ vs. $P/P_{th}$. The dash-dotted line
    indicates the minimum uncertainly of $\sigma_k \cdot \sigma_{r}=1/2$. }
    \label{fig:spotsize}
\end{figure}

Due to the discrete jump in quantum efficiency at the condensation
threshold (Fig.~1 in Ref.~\cite{Deng_PNAS03}), the emission in a
condensate region is much brighter. Since the pump beam has a
Gaussian spatial profile, the center of the spot reaches a threshold
first, leading to a sharp decrease of the emission spot size at
$P_{th}$, for both a LP condensation and a photon laser. At
$P>P_{th}$, the emission spot size measures the area which reaches
the threshold. In a photon laser, threshold density is determined by
the local density of electron-hole pairs independent of the system
size. Hence the $1/e$ spot size $\omega$ can be estimated as:
\begin{eqnarray}
\omega(P/P_{th}) =\omega_p\sqrt{1-\log_2(1+\frac{P_{th}}{P})},
\label{eq:omega_ph}
\end{eqnarray}
where $\omega_p$ is the pump spot size. Eq.~\ref{eq:omega_ph}
describes very well the photon laser data (Fig.~6C in
Ref.~\cite{Deng_PNAS03}), but fails to explain the data of a LP
condensate (Ref. \cite{Deng_PNAS03} and Fig.~\ref{fig:spotsize}A).
Here we propose that for a LP condensate, $\omega(P/P_{th})$
reflects the size of the condensate in which a condensation
threshold is satisfied. Then the critical LP density $n_c(\omega)$
increases with the system size $\omega$, and the pumping rate
$P/P_{th}$ needs to be modified as
$\frac{P}{P_{th}}\frac{n_c(\omega_c)}{n_c(\omega(P/P_{th}))}$, where
$\omega_c$ is the condensate size at $P=P_{th}$. As a simplified
model, consider $n_c$ for a 2D boson gas confined in a finite size
$L=2\omega$ \cite{Ketterle1996a}: $n_c(\omega) =
\frac{2}{\Lambda_T^2}\ln (\frac{2\omega}{\Lambda_T})$, then we
obtain:
\begin{eqnarray}
\omega(P/P_{th}) =
\omega_p\sqrt{1-\log_2(1+\frac{P_{th}}{P}\frac{\ln(2\omega/\Lambda_T)}{\ln(2\omega_c/\Lambda_T)})}.
\label{eq:omega_pol2}
\end{eqnarray}
Eq.~\ref{eq:omega_pol2} fits the data very well (solid line in
Fig.~\ref{fig:spotsize}A), with $\omega_p$ and $\omega_c$ as fitting
parameters. This suggests that $2\omega$ measures the size of the
coherent condensate above threshold.

With the distribution functions in both spatial and momentum
domains, we can evaluate how well the system can be described by a
single-particle wavefunction. The standard deviation $\sigma_k$ and
$\sigma_r$ are calculated from the momentum and spatial distribution
data $f(k)$ and $f(r)$ at $P>P_{th}$, respectively; their product is
compared to the Heisenberg minimum uncertainty limit in
Fig.~\ref{fig:spotsize}B. The sharp decrease of
$\sigma_k\cdot\sigma_r$ at $P\sim P_{th}$ indicates that a large
number of the LPs in the system condense into a single quantum
state. Deviation from the Heisenberg limit shows that there are some
thermal LPs coexisting with the coherent condensate. The slight
increase of $\sigma_k\cdot\sigma_r$ at $P/P_{th}>1$ may be caused by
condensate depletion due to LP-LP interactions at high densities
\cite{Sarchi_arxiv06,Orzel2001a_Kasevich}.

In conclusion, we studied the spatial coherence of a microcavity
polariton condensate. A Young's double-slit setup is implemented to
measure the first order coherence function $g^{(1)}(r)$ of the LPs.
The system acquires macroscopic coherence above a condensation
threshold, manifested as a sudden jump of $g^{(1)}(r)$. The observed
$g^{(1)}(r)$ vs. $r$ is well described by the Fourier transform of a
degenerate Bose-Einstein distribution in the momentum space. The
coherent condensate expands from the central region of the pump spot
to the full pump spot size, and a slow growth of the condensate size
is well understood by a simple model of quasi-BEC with a finite
size. We also confirm the position and momentum uncertainty product
$\sigma_k\sigma_r$ of the LPs decreases toward the Heisenberg
uncertainty limit above a condensation threshold.

\begin{acknowledgments}
The authors thank C.W. Lai and N.Y. Kim for useful comments, and
acknowledge support by the Quantum Entanglement Project, SORST, JST
and Special Coordination Funds for Promoting Science and Technology.
\end{acknowledgments}


\begin{thebibliography}{30}
\bibitem[{not({\natexlab{b}})}]{note:threshold}
\bibinfo{note}{The operational threshold is defined at the steepest slope of
  the input-output curve for LPs at $k=0$.}
\bibitem[{Lai()}]{Lai_spotsize}
\bibinfo{note}{Chih-Wei Lai et.al., to be publisehd.}
\end{thebibliography}
\end{document}